\documentclass[a4paper,fleqn]{cas-dc}

\usepackage[numbers]{natbib}

\def\tsc#1{\csdef{#1}{\textsc{\lowercase{#1}}\xspace}}
\tsc{WGM}
\tsc{QE}

\usepackage[ruled, vlined]{algorithm2e}
\usepackage{algorithmic}
\usepackage{amsmath,amsfonts}   
\usepackage{tabu}                      
\usepackage{booktabs}                  
\usepackage{lipsum}                    
\usepackage{mwe}                       
\usepackage{placeins}

\usepackage{cleveref}
\crefname{figure}{fig}{figures}
\Crefname{figure}{Fig}{Figures}

\usepackage{graphicx}
\usepackage{subfigure}
\usepackage{caption2}
\usepackage{float}
\usepackage{lineno}

\begin{document}




\newcommand{\ie}{i.e.}
\newcommand{\eg}{e.g.}
\newcommand{\etal}{et al.}
\newcommand{\N}{\mathcal{N}}
\newcommand{\D}{\hat{\mathcal{D}}}
\newcommand{\F}{\hat{\mathcal{F}}}
\newcommand{\R}{\hat{\mathcal{R}}}

\definecolor{green}{rgb}{0, 0.5, 0}
\definecolor{orange}{rgb}{0.6, 0.3, 0.1}
\definecolor{red}{rgb}{1.0, 0.0, 0.0}
\definecolor{teal}{rgb}{0.0, 0.4, 0.4}
\definecolor{purple}{rgb}{0.65,0,0.65}
\definecolor{saffron}{rgb}{0.95,0.75,0.2}
\definecolor{turquoise}{rgb}{0.0,0.5,0.5}
\definecolor{brown}{rgb}{0.5, 0.16, 0.16}
\definecolor{brickred}{rgb}{.6, .2 .1}
\definecolor{coral}{rgb}{1,0.45,0.33}
\definecolor{newcolor}{rgb}{.8,.349,.1}
\definecolor{mygreen}{RGB}{15, 153, 5}
\definecolor{myorange}{RGB}{255, 153, 5}

\newcommand{\Yuc}[1]{{\color{blue} #1}}
\newcommand{\yu}[1]{{\color{green} #1}}

\newcommand{\dc}[1]{{\color{red} #1}}
\newcommand{\dcc}[1]{{\color{red}[DC: #1]}}



\let\WriteBookmarks\relax
\def\floatpagepagefraction{1}
\def\textpagefraction{.001}

\shorttitle{<TWVis>}    

\shortauthors{<Siyi Li et al.>}  

\title [mode = title]{AI-based experts’ knowledge visualization of cultural heritage: A case study of Terracotta Warriors}  



%

\author[1]{Siyi Li}\fnref{fn1}
\author[1]{Yue Jiang}\fnref{fn1}
\author[2]{Bowen Jing}\corref{cor1}
\author[1]{Liuyuxin Yang}
\author[1]{Yuhe Zhang}\corref{cor1}

\cortext[cor1]{Bowen Jing and Yuhe Zhang are the corresponding authors, emails: jing\_bowen@hotmail.com, zhangyuhe0601@nwu.edu.cn}
\fntext[fn1]{These authors contributed equally: Siyi Li, Yue Jiang.}


\affiliation[1]{organization={School of Information Science and Technology, Northwest University},
            addressline={Xuefu Road, No. 1, Changan District}, 
            city={Xi’an City},
            postcode={710127}, 
            state={Shaanxi Province},
            country={China}}

\affiliation[2]{organization={Emperor Qinshihuang's Mausoleum Site Museum},
            addressline={Qinling North Road, Lintong District}, 
            city={Xi’an City},
            postcode={710699}, 
            state={Shaanxi Province},
            country={China}}



\begin{abstract}
Advancements in 3D modeling, digital display technologies, and the growing availability of digital cultural heritage data have significantly improved the accuracy of heritage depictions and expanded opportunities for analysis. However, while many studies focus on presenting specific cultural heritage figurines, an often overlooked aspect is the visualization of the Terracotta Warriors as a unified entity. This involves concisely representing the distribution of features and their relationships, providing a clear and insightful presentation that engages practitioners, academics, and wider audiences. To tackle the challenges mentioned above, this research seeks to explore the application of AI methods in processing cultural heritage data. It aims to optimize and augment the dataset, analyze the distribution and relationships of various attributes, and interpret the analysis results through visualization techniques. The Terracotta Warriors, among China's most significant cultural heritages and renowned for their abundance, exquisite workmanship, and magnitude, are chosen as a case study. The contribution of this paper is primarily twofold. Firstly, we constructed a dataset of Terracotta Warriors from Pit No. 1, detailing the attributes significant for identifying different Terracotta Warriors. Secondly, we employ various AI methods, such as generative adversarial network(GAN) and random forest, to process and analyze these attributes, followed by visualizing the analysis results for an intuitive presentation. This study introduces a novel scheme for presenting information on a collection of cultural relics, offering a practical case for analyzing and visualizing the Terracotta Warriors' attributes as a whole entity, rather than showcasing individual relics' information in isolation. This approach can be applied to other cultural heritage artifacts, such as Chinese ceramics, and supports further multidisciplinary research, including AI-based cultural heritage analysis, processing, and presentation.



\end{abstract}




\maketitle


\section{Introduction}
The rapid advancements in 3D modeling and digital display technologies, such as scanner devices \cite{BRUNO201042}, 3D modeling \cite{scopigno20113d}, and virtual reality (VR) techniques \cite{10.1007/978-3-319-91125-0_35, ma2023research}, have enabled highly accurate depictions of cultural heritage \cite{Díaz-Marín:2015:Virtual}.  Additionally, the growing availability of digital cultural heritage data has introduced new avenues for analysis and expanded access for cultural practitioners, academics, and heritage visitors alike \cite{Windhager:2019:VCH}. As a result, various forms of heritage presentation have been significantly advanced.

However, while many of these studies have concentrated on presenting cultural heritage in its entirety, often accompanied by extensive textual descriptions, an essential aspect has been overlooked: heritage narratives, this involves concisely visualizing heritage's rich features and underlying information, rather than focusing solely on its outward appearance. Thus, it is essential to first process the data, then extract key information, and finally present it in a manner that offers a clear and insightful understanding of cultural heritage. This approach not only aids cultural practitioners and academics in engaging with heritage on a broader scale but also empowers both current and potential visitors, enriching their experience and fostering a deeper appreciation of cultural heritage.



Terracotta Warriors (TW), which is a collection of thousands of life-sized sculptures buried with China's first emperor, Qin Shi Huang\cite{Nickel:2013:First}, serves as our case study. TW is regarded as one of the greatest archaeological discoveries of the 20th century, offering profound insights into ancient Chinese culture and history. The detailed attributes of the TW, such as their hairstyle and headgear styles, also provide valuable information for studying the Qin dynasty's complex military system, social structure, and cultural arts as well.

Artificial Intelligence (AI), renowned for its efficiency and effectiveness, has attracted considerable interest from researchers across various fields, and cultural heritage is no exception \cite{TOWAREK202464}. AI methods can address many challenges associated with large-scale data analysis and offer the potential for deeper interpretation, revealing and presenting relationships between attributes that were previously difficult to detect.

Therefore, in this study, we first applied various AI techniques to process data from the Terracotta Warriors of Pit No. 1, and then we analyzed the significance and distribution of each attribute, uncovered the relationships between them, and presented the results through a range of visualization methods. Finally, we conducted a user study, gathering evaluations and suggestions from experts in the field, including Terracotta Warriors scholars, graduate students in cultural heritage, and history professors, which underscored the usability and benefits of our work.


\section{Research Aims}

This research aims to investigate the use of AI techniques in processing cultural heritage data, with a focus on interpreting and conveying key information. The Terracotta Warriors (TW) serve as the case study for this exploration. Specifically, we first developed the TW Dataset, TW-1087S, by augmenting the original TW-1087 dataset using a GAN-based method. The TW-1087 dataset comprises attributes of 1,087 Terracotta Warriors from Pit No. 1, but it suffers from significant class imbalances and missing data. We then focused on analyzing the dataset of the TW by applying several AI methods, for example, using the Random Forest(RF) method to compute the feature importance for TW classification,
and employing the Cramér’s V coefficient for computing the correlations between the attributes. The analysis results are finally visualized in a more structured form to convey their meanings and heritage narratives, providing an easier way for academics, students, and history enthusiasts to quickly learn about the attributes of the TW.

\section{Related Work}

We review the visual analysis methods for cultural heritage, multi-view visualization, hierarchical visualization, and the visualization methods for multivariate data.

\subsection{Visualization of Cultural Heritage}
The conservation of cultural heritage has become a crucial aspect of modern society, and the increase in digital cultural heritage data has provided new modes of analysis and higher levels of access for both domain experts and general users\cite{Windhager:2019:VCH}. The swift advancements in 3D technology have facilitated an increasingly precise depiction of our cultural heritage\cite{Díaz-Marín:2015:Virtual}. The visualization of cultural heritage is commonly achieved through various methods such as 3D visualization\cite{Mass:2001:3DDCH, Hu:2022:Measurement}, 3D modeling\cite{Koller:2010:RCDA3DCHM, Pavlidis:2007:methods}, augmented reality\cite{Wang:2018:AR}, and virtual reality\cite{Guido:2019:IVFUIECH}. These advanced technologies are used to create immersive and interactive experiences for visitors\cite{Díaz-Marín:2015:Virtual}, allowing them to explore cultural heritage sites and artifacts in unprecedented detail and depth, offering enormous potential for the conservation and presentation of our invaluable cultural heritage\cite{Ruecker：2011:Visual}.  However, while most of these studies have attempted to present cultural heritage as it is, they can also easily overwhelm visitors due to the wealth of perception and information associated with the cultural artifacts\cite{Stephen:2009:MF}. TWVis aims to use visualization methods that combine artifacts characteristics with on-site artifacts entities to improve the visitor experience, giving them a deeper sense of the cultural context of the heritage.

\subsection{Multi-View Visualization}
Numerous studies on visualization have advanced our understanding of the design space for multi-view visualizations and their effectiveness for various types of visualizations, tasks, and datasets\cite{Shaikh:2022:MV}. These studies have provided valuable insights into the field of visualization. There are numerous papers investigating various design options for creating multi-view visualisations\cite{Michael:2011:VCIV, Javed:2012:EDSCV, LYi:2021:CLR}. Inspired by Gleicher \textit{et al.}\cite{Michael:2011:VCIV} which designed three basic building blocks for arranging multiple views to support comparison tasks and Javed \textit{et al.}\cite{Javed:2012:EDSCV} that introduced four operations to combine multiple views, L' Yi \textit{et al.}\cite{LYi:2021:CLR} further explored these arrangement techniques in a study that contributed to the expansion of the design space. Additionally, several studies have focused on providing design guidelines based on empirical studies\cite{Liu:2015:ERJGPMV, Ondov:2019:FF, Qu:2018:KMVC, Jardine:2020:PPVC}. Moreover, researchers have conducted many controlled user studies to evaluate the usefulness of arrangement types for selected tasks\cite{Jardine:2020:PPVC, Cho:2016:VAiRoma, Wagner:2019:KAVAGait, Han:2022:HisVA}. Qu \textit{et al.}\cite{Qu:2018:KMVC} conducted a study namely Wizard-of-Oz to understand how visual channels can be used consistently in multiple views. Han \textit{et al.} \cite{Han:2022:HisVA} proposed HisVA, which is an effective multi-view exploration space that evaluates systems using a user study with qualitative analysis of user exploration strategies. The VAiRoma proposed by Cho \textit{et al.}\cite{Cho:2016:VAiRoma} used user studies to validate that the arrangement of the system could enable participants to discover new findings beyond those they had gleaned from books and web searches. 

In this paper, we use the design guidelines advocated in the field of multi-view visualization to identify solutions to the prevalent multi-view usability problems we describe.

\subsection{Hierarchical Visualization}
There is a wide range of available visualization methods to help comprehend hierarchical data patterns and distributions, including Icicle Plot\cite{Kruskal:1983:Icicle}, Treemap\cite{Johnson:1998:Tree}, and Sunburst\cite{Stasko:2000:Focus}. Stasko \textit{et al.}\cite{Stasko:2000:Focus} compared Sunburst and Treemap in a document search task and verified that Sunburst is a more efficient and clearer presentation of the hierarchy. Hierarchical data is ubiquitous in real-life scenarios, and researchers have explored various ways to integrate Sunburst diagrams with different fields to provide novel insights and designs. Such integration has led to further optimization and improvement of the Sunburst diagrams. Rodden \textit{et al.}\cite{Rodden:2014:ASV} utilized sunburst diagrams to analyze user paths on websites. Liu \textit{et al.}\cite{Liu:2015:POA} proposed the Necklace Sunburst algorithm to address the issue of insufficient display of leaf nodes in radial loops, and applied it to a large-scale public opinion analysis system with promising results. Furthermore, Médoc \textit{et al.}\cite{Médoc:2019:UMM} employed Sunburst visualizations to aid analysts in exploring social media messages for anticipating emergency responses following disasters.

\subsection{Visualization of Multivariate Data} 

Over the past twenty years, a variety of new methods for visualizing multivariate data have emerged, and some efforts have been made to survey these approaches \cite{Liu:2017:VHDD, Zhou:2019:Multivariate, Nobre:2019:Lineage}. Parallel coordinates is an efficient visualization method\cite{Inselberg:1990:PC}, often used in the visualization of high-dimensional geometry and multivariate data\cite{Viau:2010:FPSM}. Tyagi \textit{et al.}\cite{Tyagi:2023:PC-Expo} introduced PC-Expo, an interactive method based on metrics for reordering axes in parallel coordinate displays, which provides an effective means of optimizing the display of multivariate data. 

 A multivariate network is a graph in which nodes and edges are endowed with a diverse array of attributes, potentially conveying nuanced information and yielding rich insights\cite{Fung:2016:Person}. The system proposed by Lex \textit{et al.}\cite{Lex:2010:Caleydo} employs linking and brushing to establish connections between nodes and their respective attributes, revealing the relationship between topology and attributes through interaction. Several layout techniques propose aggregated, simplified layout solutions for nodes or edges to solve the visual clutter caused by overlapping edges. Lhuillier \textit{et al.}\cite{Lhuillier:2017:FFTEB} improved on existing edge binding methods to implement selective edge bundling based on edge attributes and for large datasets. Bouts \textit{et al.}\cite{Bouts:2015:CER} proposed a new edge clustering routing algorithm that allows user-defined edge clustering. In our visualization of the characteristics of the TW of the First Emperor of Qin, we have focused on visualizing the multiple attributes of the data entities and the relational attribute data between the entities to help users quickly obtain information on the unique characteristics of the TW of different types.

\section{The TW-1087S Dataset}
According to "The Excavation Report of Terracotta
Warriors and Horses Pit No.1 of Qin Shihuang’s Mausoleum
1974-1984 (Volumes I \& II)", we have developed the Terracotta Warriors Dataset TW-1087, which contains attributes of 1087 terracotta warriors in Pit No.1, including the type and size of their hairstyle, armor, and headgear, among others.

\subsection{Details of Dataset TW-1087}

\begin{table*}[b]
  \caption{%
   The detailed values of each attribute in TW-1087.%
  }
  \label{tab1}
  \scriptsize%
  \centering%
 \resizebox{\textwidth}{!}{
  \begin{tabu}{%
     l%
      *{6}{l}%
      *{5}{l}%
   }
   \toprule
   Attributes & Values & Definition \\
   \midrule
   c\_id & $[1,11]$ or K & The $ith$ corridor. \\
  \midrule
  t\_id & 1, 2, 10, 19, 20 & The $ith$ trench.\\
  \midrule
  crops & 0 (chariot soldier), 1 (infantryman) & The army class to which the TW belongs. \\
  \midrule
      position & 0 (following vehicles), 1 (independent), 2 (onboard) & The military formation’s location where the TW is situated.\\
  \midrule
  height & float values & The height(foot to head) of the TW. \\
  \midrule
  weapon  & 0 (archery), 1 (long weapons), 2 (swords), 3 (none) & The weapons that the TW held.  \\
  \midrule
  hairstyle & 0 (cone bun), 1 (flat bun)& The hairstyle which the TW held. \\
   \midrule
  headgear & 0 (double-plate crown), 1 (single-plate crown), 2 (He crown), 3 (none), 4 (hood) & The headgear which the TW held. \\
  \midrule
  robe\_num & 1, 2 & The robe layers number which the TW dressed. \\
  \midrule
  armor\_type & 0 (type III B), 1 (none), 2 (type I A), 3 (type II A), 4 (type II B), 5 (type I B), 6 (type III A) & The armor type which the TW dressed. \\
  \midrule
  tw\_class & categorical & The class to which the TW belongs. \\

    \bottomrule 
  \end{tabu}%
  }
\end{table*}

The TW-1087 dataset comprehensively documents the information of 1087 terracotta warriors in Pit No. 1 at Emperor Qin Shi Huang's Mausoleum, detailing 10 feature attributes and one class label, as listed in Table \ref{tab1}. It is noteworthy that among these 10 attributes, there is a numeric attribute (\textit{i.e.}, height) and 9 categorical attributes. 

These 1,087 figurines can be categorized in several ways: by attire, they fall into robed and armored figurines; by rank, into military officials and ordinary soldiers, with military officials further classified into high, middle, and low ranks. Based on combat function, they are divided into vehicle soldiers and infantry soldiers, with vehicle soldiers including specialized roles such as charioteers and chariot soldiers. Based on the classification criteria outlined in the archaeological report, the 1,087 terracotta figurines are grouped into seven distinct classes: 396 Robed Warriors (RW), 633 Armored Warriors (AW), 8 Charioteers (CS),  8 Chariot Soldiers on Right (CT), 5 High-Ranking Officials (HR), 10 Middle-Ranking Officials (MR), and 27 Low-Ranking Officials (LR).


Since the data reported in the excavation report is handwritten records of archaeologists, 567 entries, occupying 52.16\% of the total, were found to have missing data. Directly deleting these entries would severely affect the integrity of the data, therefore, this paper first adopted the GAIN imputation(Generative Adversarial Imputation Nets) \cite{yoon2018gain} algorithm to fill in missing values. 


Furthermore, the TW-1087 dataset displays a significant scarcity of minority classes, leading to considerable class imbalance issues. Statistics indicate that the AW class constitutes 58.23\% of the total, while the HR class, the smallest, accounts for merely 0.46\%. This imbalance significantly limits the model's learning performance, especially for minority classes. The model often neglects these classes during training, leading to high accuracy for the majority but poor results for the minority. As a result, it hampers the identification of key features and obstructs deeper insights into the minority classes. Therefore, we also augment the dataset before proceeding with our analysis, forming a new dataset named TW-1087S.

   

\subsection{Missing Data Imputation}
Given the TW-1087 dataset, we first adopted the GAIN imputation\cite{yoon2018gain} algorithm to fill in missing values in the TW-1087 dataset.

GAIN \cite{yoon2018gain} is a missing data imputation algorithm built on generative adversarial networks. Its key innovation lies in using a generator to predict missing values while incorporating a hint vector as auxiliary information. The hint vector, combined with the predicted values, is input into the discriminator, which assesses the difference between the predicted and actual values, continuously refining the network's parameters. This approach enables GAIN to progressively approximate the true data distribution, ultimately providing accurate imputation of missing data.

Since GAIN\cite{yoon2018gain} requires a complete dataset for benchmark comparisons, we selected 520 complete samples from the TW-1087 dataset to form the TW-520 dataset, which serves as the experimental dataset. Missing data in the original TW-1087 dataset primarily affects the "hairstyle," "headgear," "weapon," and "height" attributes. To simulate missing entries, 30\% of the data from these four features is deliberately removed. A random seed is employed to ensure reproducibility and consistency, minimizing the influence of randomness on the results. The modified TW-520 dataset, now containing missing values, is then used to train the imputation model, with the results compared against the original TW-520 dataset to evaluate imputation accuracy. 

To validate the imputation performance of the GAIN\cite{yoon2018gain} on the TW-520 dataset, we compare the results of GAIN with that of four imputation algorithms, including traditional statistical analysis imputation (STA), Multiple Imputation by Chained Equations (MICE)\cite{MICE}, SGAIN \cite{SGAIN}, and WSGAIN\cite{WSGAIN}. The differences in average values of accuracy, F1-score, and AUC for the five classifiers—including Logistic Regression (LR), Decision Tree (DT), Random Forest (RF), Multi-layer Perceptron (MLP), and Support Vector Machines (SVM)—are utilized as evaluation metrics. Table \ref{imputationResults} lists the scores. It can be observed that GAIN \cite{yoon2018gain} outperforms the other baselines, achieving the best difference scores.

\begin{table}[b]
  \caption{%
   The differences in classification performance on the TW-520 dataset following the imputation of missing data. Best scores are in \textbf{bold}.}
  \label{imputationResults}
  \scriptsize%
  \centering%
  \begin{tabu}{lccc}
   
   \toprule
    Methods & Avg Accuracy & Avg F1-score & Avg AUC\\   
   \midrule
    STA & 1.154 & 0.014 & 0.003\\
    MICE\cite{MICE} & 1.923 & 0.146 & 0.007\\
    SGAIN\cite{SGAIN} & 0.385 & 0.005 & 0.001\\
    WSGAIN\cite{WSGAIN} & 1.154 & 0.015 & 0.003\\
    GAIN\cite{yoon2018gain} & \textbf{0.384} & \textbf{0.004} & \textbf{0.000}\\
  \bottomrule           
  \end{tabu}%
\end{table}

\subsection{Data Augmentation}
The smallest minority class, the HR class, contains only five samples, making it challenging to apply generative data augmentation methods like Table-GAN \cite{2018Data}. Table-GAN\cite{2018Data} requires sufficient samples to accurately learn the data distribution, particularly for minority classes, which results in its failure on the TW-1087 dataset. Therefore, we employ a novel two-stage data augmentation strategy that combines the oversampling algorithm SMOTENC\cite{chawla2002smote} with a proposed generative approach Table-CGAN.

\subsubsection{Data augmentation methods}
The SMOTENC \cite{chawla2002smote} algorithm is employed to generate new instances from the existing samples. SMOTENC \cite{chawla2002smote} is specifically designed to address the issue of minority class samples in imbalanced datasets. It increases the number of minority class samples by inserting new synthetic instances between existing minority samples, thereby aiming to balance the class distribution. This approach not only mitigates underfitting in Table-GAN\cite{2018Data} caused by the sparse minority class samples but also ensures that the generator produces genuinely new samples rather than simply replicating existing ones. 

While SMOTENC effectively reduces class imbalance by interpolating between minority class samples to generate new ones, it may limit the diversity of these samples, often concentrating them within the existing minority class feature space. Therefore, we propose a novel method named Table-CGAN, which is capable of learning the underlying distribution of the data and generating new, diverse synthetic samples, helping to further enrich the dataset's feature representation and enhancing the model's ability to generalize to unseen data. Table-CGAN consists of three components—a generator, a discriminator, and a classifier—and utilizes generative adversarial networks (GANs) to synthesize fake tables that are statistically similar to the original table while preventing information leakage. It integrates the principles of CGAN\cite{CGAN} with the foundational framework of Table GAN\cite{2018Data}, facilitating the generation of new samples for minority classes in tabular data.

By applying the above two-stage data augmentation strategy, a new dataset, TW-1087S, comprising 1,800 samples, is created. The TW-1087S dataset will be the focus of the following attribute analysis. 

\subsubsection{Data augmentation results evaluation}
To evaluate the data augmentation results, we apply a Random Forest classifier to the TW-1087 and TW-1087S datasets, respectively. The classification performance on the TW-1087 and TW-1087S datasets are compared, and three evaluation metrics are used to assess the effectiveness of the data augmentation. For these metrics, higher scores indicate better performance. The specific scores are presented in Table \ref{tab4}.


   
            


\begin{table}[b]
  \caption{%
   The classification performance on TW-1087 and TW-1087S. Pre is precision, Re is recall and F1 stands for F1 score.%
  }
  \label{tab4}
  \scriptsize%
  \centering%
  \begin{tabu}{%
     c%
      *{6}{c}%
      *{5}{c}%
   }
   
   \toprule
            \multirow{3}{*}{Label} &
            \multicolumn{4}{c}{TW-1087} & \multicolumn{4}{c}{TW-1087S} \\
            
            \cmidrule(r){2-5} \cmidrule(r){6-9}
            & Pre & Re & F1 & AUC & Pre & Re & F1 &  AUC \\
            \midrule
            RW & 0.99 & 0.99 & 0.99 & & 0.99 & 0.99 & 0.99 & \\
            AW & 0.94 & 0.99 & 0.97 & & \textbf{0.95} & 0.99 & 0.96 & \\
            CT & 0.99 & 0.99 & 0.99 & & 0.99 & 0.95 & 0.92 & \\
            CS & 0.99 & 0.50 & 0.67 & 0.85 & 0.99 & \textbf{0.95} & \textbf{0.97} & \textbf{0.98}\\
            LR & 0.75 & 0.60 & 0.67 & & \textbf{0.94} & \textbf{0.85} & \textbf{0.89} & \\
            MR & 0.99 & 0.33 & 0.50 & & 0.94 & \textbf{0.85} & \textbf{0.89} & \\
            HR & 0.00 & 0.00 & 0.00 & & \textbf{0.99} & \textbf{0.90} & \textbf{0.97} & \\

  \bottomrule           
  \end{tabu}%
\end{table}


The experimental results reveal that before data augmentation, the HR class within the TW-1087 dataset exhibited precision, recall, and F1 scores of zero, underscoring the model's inability to detect any samples from this minority class within the highly imbalanced dataset. Furthermore, the low recall rates for the MR, CS, and LR classes suggest the omission of numerous true positives. After data augmentation, the dataset's overall accuracy surged to 97\%, accompanied by notable enhancements in precision and recall across most classes. Additionally, the AUC metric ascended from 85\% to 98\%, implying that employing data augmentation techniques on TW-1087 not only effectively addresses class imbalance concerns but also elevates the overall dataset quality. This enhancement holds significant importance in bolstering the accuracy of classification tasks, particularly in distinguishing TW types.

\section{Attribute Analysis and Visualization}
\label{Attribute Analysis}

This section centers on the visual analysis of TW attributes, utilizing both AI-based methods and visualization technologies to systematically investigate the relationships between attributes and classes, as well as correlations within the attributes themselves. Specifically, we first assess attribute importance to isolate critical attributes, facilitating a focused visual analysis of the most influential attributes for discerning TW. Subsequently, these key attributes are visualized using box plots to illustrate their distribution and significance, revealing additional insights underlying these attributes. Meanwhile, attributes of lower importance are visualized using violin plots, highlighting their distribution and illustrating why they hold less significance for classification.

\subsection{The Importance of the Attributes}

\begin{figure*}[htbp]
    \centering
 \includegraphics[width=0.55\textwidth]{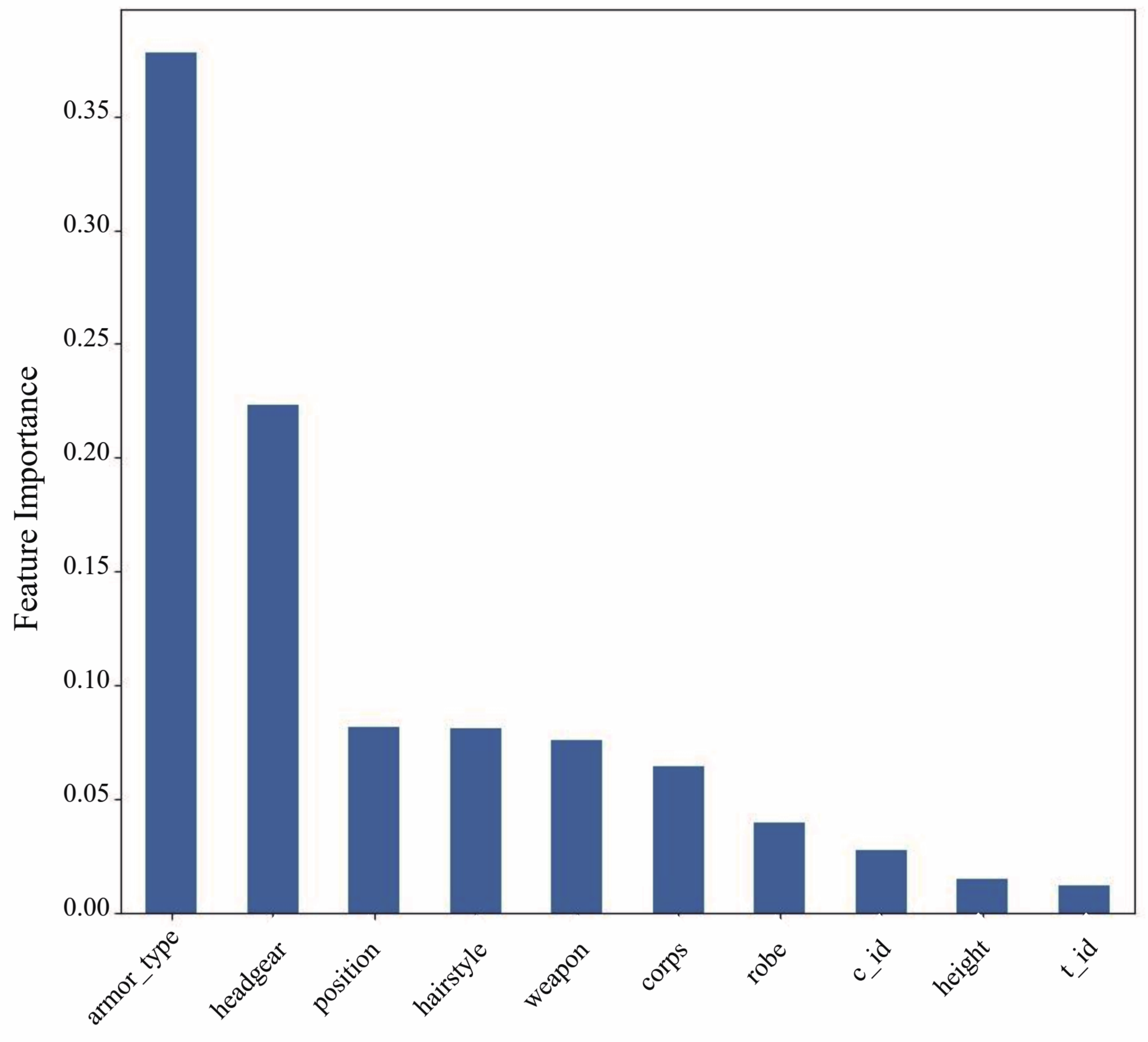}
    \caption{Ranking of Feature Importance for the TW-1087S.}
    \label{featurImportance}
 \end{figure*} 

Initially, we utilize a Random Forest(RF) classification model \cite{breiman2001random} to assess the importance of various attributes in identifying key factors for classifying TW. Random Forests consist of an ensemble of decision trees, each trained on a subset of the dataset. By aggregating the predictions of these trees through majority voting in classification tasks, the model enhances both accuracy and robustness. Furthermore, RF can be leveraged to evaluate feature importance by measuring the extent to which each feature contributes to reducing uncertainty or impurity during the decision tree construction process. In this study, we employ Gini impurity both to rank feature importance and to guide the construction of the decision trees.

The attribute importance obtained by RF is depicted in Fig.\ref{featurImportance}. It can be seen that attributes such as "armor\_type" and "headgear" exert a significant influence on the model's predictive performance. Conversely, attributes like "c\_id", "height", and "t\_id" demonstrate lower importance and offer a minimal predictive contribution. This finding lays the groundwork for identifying the key attributes crucial for classifying TW. Based on the analysis, it reveals that using armor is a more feasible approach for classifying the Terracotta Warriors into more categories. This underscores the significance of our work: the TWs possess numerous visual features, and through visual analysis, we can uncover potential insights. For example, headgear is mainly used for determining TWs' ranks, however, for multidimensional classifications, armor appears to more accurately reflect the figures' categories than headgear.

\subsection{Attribute-class Analysis and Visualization}
\label{boxplots}

Given the key attributes of TW, we then focus on a concentrated analysis of these key attributes and their intrinsic connections with TW classes, exploring how these relationships reflect the historical and cultural values behind the TW. The aim is to deepen the comprehensive understanding of TW attributes, thereby increasing the value of this dataset in academic and applied fields.

\begin{figure*}[htbp]
    \centering
    \includegraphics[width=0.55\textwidth]{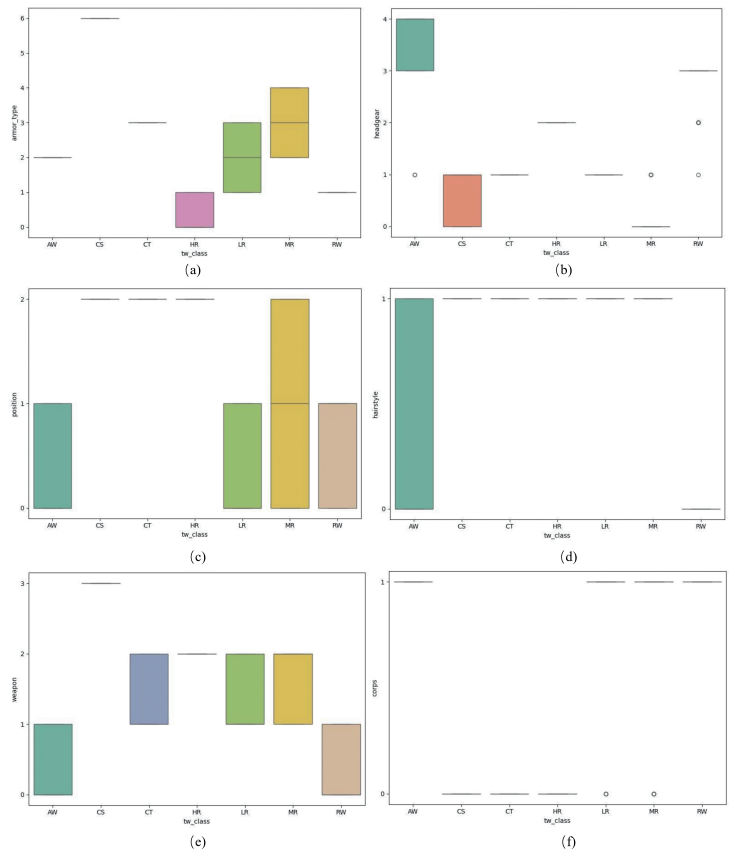}
    \caption{Boxplots of the top six attributes in ranked importance in the TW-1087S.}
    \label{4boxFeature}
\end{figure*} 


Box plots are a powerful tool for showing data distributions and statistics, providing a concise view of the central tendency and spread of the data across the range of a variable. The boxes represent attributes with multiple values, the lines correspond to individual values, and the small circles indicate outliers introduced during the data augmentation process. 

The boxplots of the top six key attributes are presented in Fig.\ref{4boxFeature}. It can be found that these six key attributes all show relatively distinctive distribution patterns in the box plots. Specifically, Fig.\ref{4boxFeature}(a) shows that "MR" is equipped with type II B, type II A, and type I A armor, while the "LR" wears either no armor or type II A and type I A armor, and the "HR" wears no armor or type III B. The other classes are each equipped with a single type of armor. According to archaeological research, type A armor has larger plates, probably representing leather armor. In contrast, Type B armor, especially II and III, has smaller plates, which probably represent the metal armor. Considering the metallurgical conditions of the Qin Dynasty, metal armor was a scarce piece of combat equipment. Thus, the differences in the "armor type" of TWs can highlight the military ranks and functions, suggesting a structured reflection of the Qin dynasty’s military organization and social hierarchy. Fig.\ref{4boxFeature}(b) illustrates the distribution of "headgear," indicating that the "CS" class and the "AW" class wear two types of headgear, while the other classes each display a specific type of headgear. This corresponds to the scenario where lower-ranking military officials wear a Single-plate Crown, middle-ranking military officials usually wear a Double-plate Crown, and high-ranking officials wear a He Crown. Fig.\ref{4boxFeature}(c) presents the boxplot for the “position” attribute, showing that “AW,” “LR,” “RW,” and most “MR” figures are either independent or positioned behind vehicles, while other classes are associated with official vehicles. This implies that Pit No.1 represents a kind of vehicle combat squad, which reflects the main military force in China during the Qin Dynasty and beforehand. The hairstyle attribute, which shows a partial correlation with headgear, is illustrated in Fig.\ref{4boxFeature}(d). It can be seen that the classes which wear crowns all have flat buns, such as the "CS",  "CT", "HR", "MR" and "LR". The "RW" class has cone buns, while the "AW" class has both flat and cone buns. Fig.\ref{4boxFeature}(e) shows that "RW" and "AW" use archery and long weapons, while "LR", "MR" and "CT" use long weapons and swords. "CS" are on vehicles and therefore do not use weapons, and "HR" only uses swords. It can be inferred that the higher the TW’s rank, the shorter their weapon and attack range, suggesting that high-ranking officials were generally engaged in the conduct of war rather than direct combat. Fig.\ref{4boxFeature}(f) shows the corps pattern of the TWs, which exhibits a strong correlation with the distribution pattern of position.

\subsection{Attribute Distribution Analysis and Visualization}

\begin{figure*}[htbp]
    \centering
    \includegraphics[width=\textwidth]{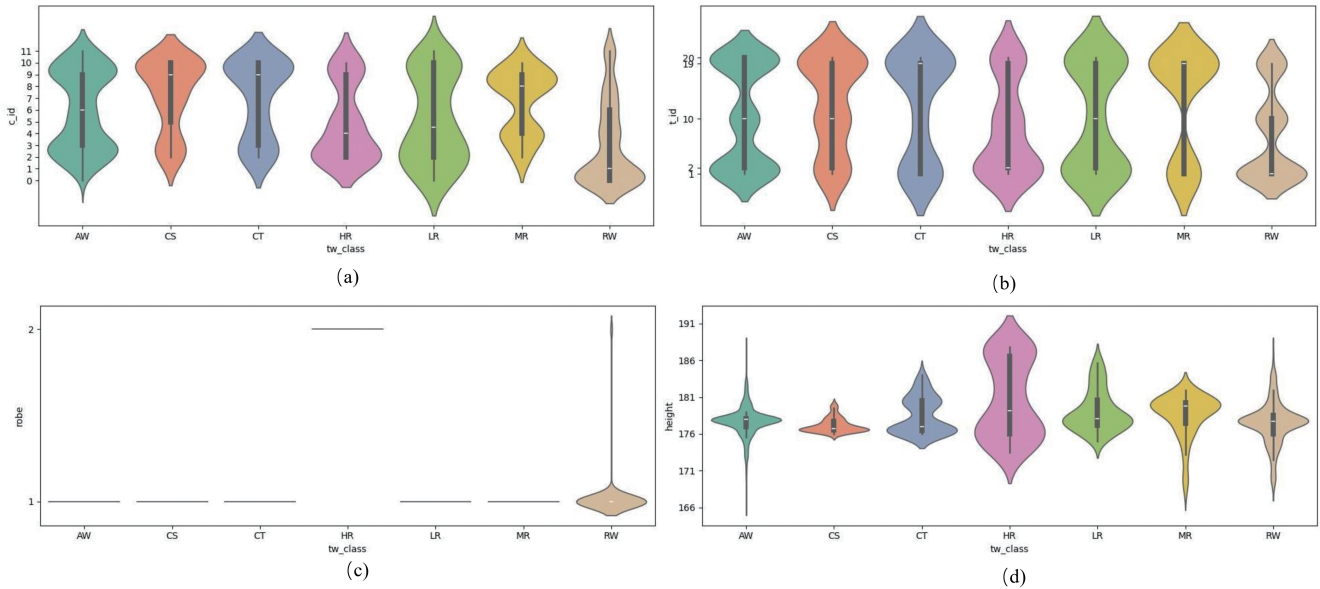}
    \caption{Violin plots of the six distinguishable attributes of TW-1087S. The x-axis delineates different classes of TW, while the y-axis illustrates the distribution of attribute values. Its width represents the data density within the numerical range; the wider it is, the denser the data.}
    \label{6ViolinsTWS}
\end{figure*}

Although some of the attributes exhibit lower importance for classification, the distribution of their values can also provide valuable insights, which can contribute to supporting heritage narratives. For instance, Fig. \ref{6ViolinsTWS} presents violin plots for the last four attributes, visualizing the distribution and probability density of values by combining features of both box plots and density plots. The thick black bar at the center indicates the interquartile range, while the thin black lines at each end represent the minimum and maximum values, highlighting the data's full range. The white dot denotes the median.

As shown in Fig. \ref{6ViolinsTWS}, despite the "c\_id" feature's lower importance, its violin plot shows a high concentration of the "RW" class near the value 0 (span), indicating that Robed Soldiers are more likely positioned at the front of Pit No. 1. The "AW" class displays a bimodal distribution at corridor \#2 and \#10, suggesting a denser presence of Armored Warriors at these corridors (A diagram of the military formation in Pit No. 1 is shown in Fig.\ref{PitNo.1} in the supplementary materials). Furthermore, as shown in Fig.\ref{PitNo.1} in the supplementary materials, Robed Warriors are positioned in the first three rows, and Armored Warriors mainly occupy corridors \#1, \#2, \#3, and \#9, \#10, \#11, which is consistent with the information shown in Fig.\ref{6ViolinsTWS}. Historical research suggests that the first three rows of figurines in Pit No.1 may have represented the vanguard of the entire force. According to historical records, Qin soldiers fought valiantly, with vanguard troops shedding their armor to charge the enemy more swiftly. This detailed mapping of distribution patterns offers a deeper understanding of ancient military formations and organizational structures, shedding new light on the composition of the Qin army and its cultural significance. Furthermore, the violin plots reveal that the "RW" class has three peak distributions at positions "trench \#1", "trench \#10", and "trench \#19", while the "AW" class peaks at "trench \#2", "trench \#10", and "trench \#20", suggesting that "RW" is likely positioned ahead of "AW". This positioning can also be corroborated by the earlier analysis of the "c\_id" feature.

Referring to the distribution of the robe attribute shown in Fig.\ref{6ViolinsTWS}(c), the "HR" class and a small subset of the "RW" class wear two layers of robes, while the remaining TWs wear only a single layer. The outer robes of the TWs generally extend to the knee or hip, designed for ease of movement and combat. According to the violin plot for height, the average height of TWs is approximately 178 cm; however, the height distribution appears scattered without a discernible pattern. This suggests that height was not a determining factor in TW classification during that era, with no indication that higher rank correlates with taller stature. According to the height attribute's violin plot, we can infer that the average height of the TW is approximately 178cm. However, there is no discernible pattern in the distribution of heights; it appears to be relatively scattered. This suggests that height was not a determining factor in the classification of TW during that era and there is no indication that higher rank correlates with taller stature.

\subsection{Attribute Correlation Analysis and Visualization}

To thoroughly explore the internal relationships between the attributes of the TW, we use Cramér's V to measure the correlations between these attributes. Cramér's V is a measure of effect size for the chi-square test of independence, used to assess the strength of the correlation between two categorical variables. Since the attribute "height" shows lower importance for classification and exhibits a less distinct distribution for each class, it is not included in this analysis. 

\begin{figure*}[htbp]
    \centering
    \includegraphics[width=0.55\textwidth]{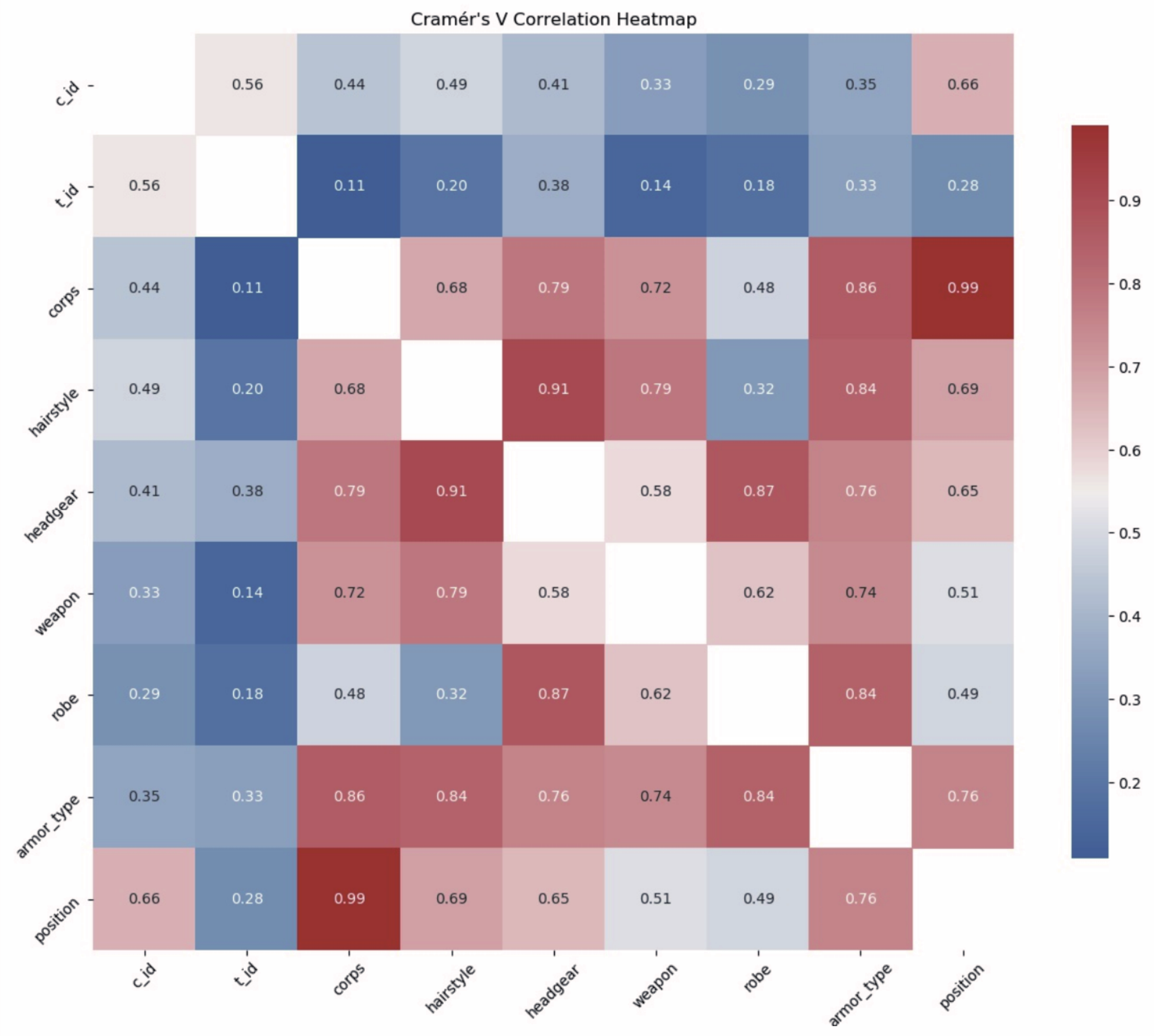}
    \caption{Attribute Correlation Matrix of the TW-1087S.}
    \label{Correlation}
\end{figure*}

The correlation results of the attributes are visually present in the form of a correlation matrix, as illustrated in Fig.\ref{Correlation}. From Fig. \ref{Correlation}, it is evident that the correlation coefficient between "corps" and "position" is 0.99, indicating a strong correlation. Actually, the position alone allows us to directly distinguish between a charioteer and an independent infantryman. The strong correlation between these two features implies a potential causal relationship. Similarly, the correlations between "corps" and "armor\_type," "headgear" and "weapon" are 0.86, 0.79, and 0.72, respectively. This suggests that different military units are distinguished by unique armor, weapons, and headgear, reflecting the decorative distinctions among various combat formations. The strong correlations among headgear, hairstyle, weapon, and armor type further indicate established decorative patterns. Additionally, the high correlation coefficient of 0.91 between "headgear" and "hairstyle" suggests a cultural or societal trend where specific styles of headgear correspond to particular hairstyles. As analyzed in Section \ref{boxplots}, this correlation may stem from practical considerations; for instance, the flat bun accommodates a crown, the cone bun suits a hood, and the bare flat and cone buns support various decorative shapes.

\section{Result: user study from domain experts}
\label{sec:UserStudy}

We conducted a well-designed interview with domain experts to demonstrate the effectiveness and usability of the proposed strategy. First, we obtained feedback from researchers specializing in the Terracotta Warriors to validate the analysis presented in Section \ref{Attribute Analysis}. Following this, we performed a quantitative evaluation to assess how well the proposed strategy aids domain experts in presenting data analysis results.

\subsection{Study design}

\textbf{Participants.} We recruited 15 domain experts (E1–E15) from three educational and research institutions in Xi'an city  to participate in our evaluation. Participants were selected based on their research backgrounds in Qin history or cultural heritage, ensuring the reliability of the feedback. Specifically, 4 experts (E1–E4) focus on research related to the Terracotta Warriors, 7 participants (E5–E11) are graduate students majoring in cultural heritage, and 4 participants (E12–E15) specialize in history. 

\textbf{Procedures.} First, we conducted one-on-one, semi-structured interviews with experts specializing in Terracotta Warriors research to validate the accuracy of the analysis presented in Section \ref{Attribute Analysis}. We present figures illustrating attribute importance (Fig. \ref{featurImportance}), boxplots of six key attributes (Fig. \ref{4boxFeature}), violin plots of four attributes with lower importance (Fig. \ref{6ViolinsTWS}), and the correlation matrix (Fig. \ref{Correlation}). Subsequently, we describe our hypotheses derived from these figures and invite the experts to evaluate their correctness. 

We then invited participants to complete a questionnaire (in Table \ref{tab:Questionnaire}) consisting of 12 questions, using a 5-point scale to assess the effectiveness and usability of the proposed analysis and visualization strategy. Before the formal study, we demonstrated how to interpret the figures by explaining all their elements, using examples that depicted the gender, major, and scores of students in a college to facilitate understanding. Participants were then given 8 minutes to explore the example figures independently.

The interview lasted approximately 30 minutes, followed by the questionnaire process, which took about 10 minutes. Throughout this period, we recorded and took notes on the entire study process.

\begin{table*}
  \caption{The questionnaire consists of three parts, \textit{i.e.}, the effectiveness (Q1-5), usability (Q6-9), and advantages (Q10-12).}
  \label{tab:Questionnaire}
  \small
  \begin{tabular}{l|l}
    \toprule
    
     Q1 & 
    The analysis results are correct. \\
    
    Q2 & 
    The utilization of key attributes in Fig. \ref{4boxFeature} are accurate and effective in distinguishing the Terracotta Warriors.\\
    
    Q3 & 
    The correlation results provide insight into the intrinsic relationships among the TW attributes.\\
    
    Q4 & 
   The figures can illustrate the distribution of the attributes.\\

    Q5  & 
    The interpretation of the figures is accurate and aligns with the information conveyed by the attributes.\\

     \midrule

   Q6  & 
     Interpreting the figures is straightforward and easy to learn. \\
    
    Q7 &  
    The proposed presentation form is helpful for domain users.\\
    
     Q8 & 
     The proposed analysis and presentation approach is more efficient.\\
    
    Q9  & 
     I intend to use the proposed presentation format to showcase my research data and apply it for other purposes.\\

    \midrule

     Q10 & 
     Using AI-based methods for data analysis is more efficient and stable. \\
    
     Q11 &  
     Employing AI-based methods for data analysis and visualizing the results can provide more inspiration.\\
    
     Q12 & 
    The analysis results provide compelling evidence to support the hypothesis.\\
    
    \bottomrule
    
  \end{tabular}
\end{table*}

\begin{figure*}[htbp]
    \centering
    \includegraphics[width=\textwidth]{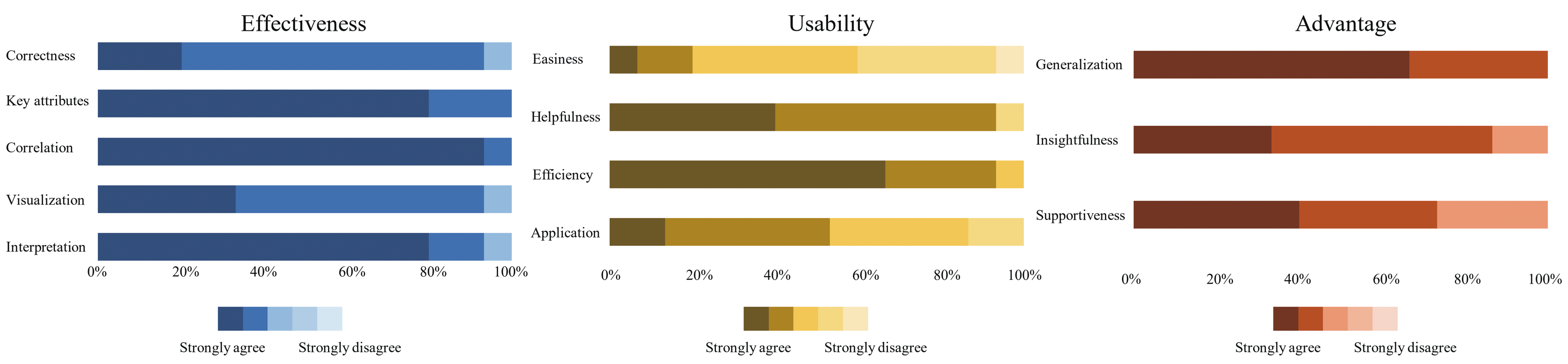}
    \caption{The summary of the feedback of the questionnaire.}
    \label{UserStudy}
\end{figure*} 

\subsection{Result}

We summarized all collected feedback on three aforementioned aspects of the evaluation in Fig. \ref{UserStudy}. 

\textbf{Effectiveness.} Most participants acknowledged the effectiveness of our method in enhancing the interpretability of TW attributes. More specifically, experts specializing in Terracotta Warriors research (E1–E4) validated the correctness of the descriptions derived from the figures in Section \ref{Attribute Analysis}. They concurred that the importance of the attributes varies, with different attributes contributing uniquely to the identification of the type of the Terracotta Warriors. Furthermore, they also agree that the "armor\_type" and the "headgear" are distinguishable for different types of Terracotta Warriors. E1 further reinforces our conclusion that "higher military ranks often wore armor composed of densely packed, narrower plates for enhanced protection." He stated: "Higher-ranking soldiers were likely equipped with metal armor, whereas lower-ranking soldiers typically wore leather armor. This distinction highlights the variations in craftsmanship and materials used for the plates, according to the soldier's rank." E3 underscores the novel insights offered by this study, particularly emphasizing the role of "armor\_type" in the multivariate classification of TWs. Experts in history (E12-E15) highly commend the fact that the distribution of attributes (Fig.\ref{4boxFeature}) can, to some extent, reflect the structure of the Qin army, offering a concise overview of its formation. The graduate students(E12-E14) also commented that the TW-1087 dataset could facilitate additional research efforts through its attribute analysis.

\textbf{Usability.} The majority of participants praised the usability of our method from both research and educational perspectives. 10/15 experts mentioned that the visualization is highly efficient, enabling them to design various presentation formats for cultural heritage or history knowledge. E1 also emphasized that attribute analysis and visualization assist in clarifying the key features when displaying the Terracotta Warriors in the museum. E3 believes that visualization can show the distributional characteristics and correlations of multiple attributes more intuitively than tables and reports, \textit{etc.}, which helps the researcher to grasp the nature of TW more easily. Additionally, from the perspective of visualization, history experts (E13-E15) believe that this analysis and visualization method serves as a powerful educational tool for teaching and disseminating historical knowledge, both in classrooms and in museum education. Five graduate students (E5–E8, E10) also believe that the proposed strategy for presenting data and information not only supports their other hypotheses but also provides additional inspiration, owing to its high efficiency in conveying and presentation of information.

\textbf{Advantages.} All Terracotta Warriors experts (E1–E4) believe that using AI-based methods for data analysis is more efficient and regard it as a novel research direction. E1 and E3 also suggested incorporating additional data into the TW-1087 dataset to enable more comprehensive and detailed research, such as the size, composition, and artistic style of each specific part of the TWs. Furthermore, the majority of history experts (E12, E14–E15) highly value the benefits of employing machine learning methods for data analysis and statistical outcomes. However, E13 suggested that highlighting specific identities within the collection would be beneficial for further analysis. From the students' perspective, all of them (E5–E11) believe that the figures and their corresponding descriptions help organize their thoughts, while the quantitative analysis provides stronger support for their hypotheses and conclusions. 

\textbf{Suggestions.} In addition to the positive feedback, a primary concern raised by the participants is the ease with which non-computer experts can generate, read, and interpret the figures. Some participants expressed that the current visualizations may be too technical, potentially alienating those without a background in computing or data analysis. Therefore, simplifying these visualizations or providing additional explanatory materials could greatly enhance comprehension and engagement among non-expert users. Furthermore, E10 also suggested incorporating interactive elements into the figures to provide laypersons with greater flexibility in quickly grasping the visualization tools.

\section{DISCUSSION AND LIMITATION}
\label{sec:discussion}

By employing and integrating various AI-based methods, this study utilizes the dataset of TW from Pit No.1 as a use case to demonstrate how experts' knowledge of cultural heritage can be interpreted and conveyed by combining AI-based methods and visualization techniques.


The major contributions of this study are as follows: (1) The development of the TW-1087S dataset, which includes 10 attributes of 1800 Terracotta Warriors from Pit No.1. This dataset encompasses detailed information on various attributes such as the type and size of their hairstyle, armor, and headgear, among others, to support the analysis of these attributes. Furthermore, we also enhance the dataset by integrating SMOTENC and Table-CGAN methods to improve the study of attribute importance and classification. (2) This study introduces a novel scheme for analyzing and presenting information about a collection of cultural relics, offering a practical case for visualizing the Terracotta Warriors, rather than showcasing a single relic in various forms. The user study involving domain experts indicates that the proposed scheme is well-received by most experts, who praise its effectiveness and high efficiency.


However, there are future improvements and challenges that need to be addressed. First, although the study uses data from Terracotta Warriors in Pit No.1, the limitation caused by the limited number of samples is inevitable despite the extensive attribute coverage. To mitigate this, the TW-1087 dataset will be enhanced by adding more data from other pits, such as Pit No.2 and Pit No.3. Second, future research could also focus on creating a hierarchical visualization that ranges from macro information (an overview of the Qin army) to micro information (an individual Terracotta Warrior) by analyzing the relationships between individuals and collections. Third, the boxplots in Fig. \ref{4boxFeature} show that a small number of outliers were introduced by the data augmentation algorithm. Although these outliers have minimal impact on the overall analysis, their introduction remains noteworthy. Avoiding the introduction of outliers or noise during data enhancement is an important area of future research.

\printcredits

\bibliographystyle{elsarticle-num}
\bibliography{template}

\bio{}
\endbio


\appendix 

\clearpage

\setcounter{figure}{0}
\renewcommand{\thefigure}{A\arabic{figure}}
\renewcommand*{\theHfigure}{\thefigure}

\section*{Supplementary Materials}

\begin{center}
    \includegraphics[width=\textwidth]{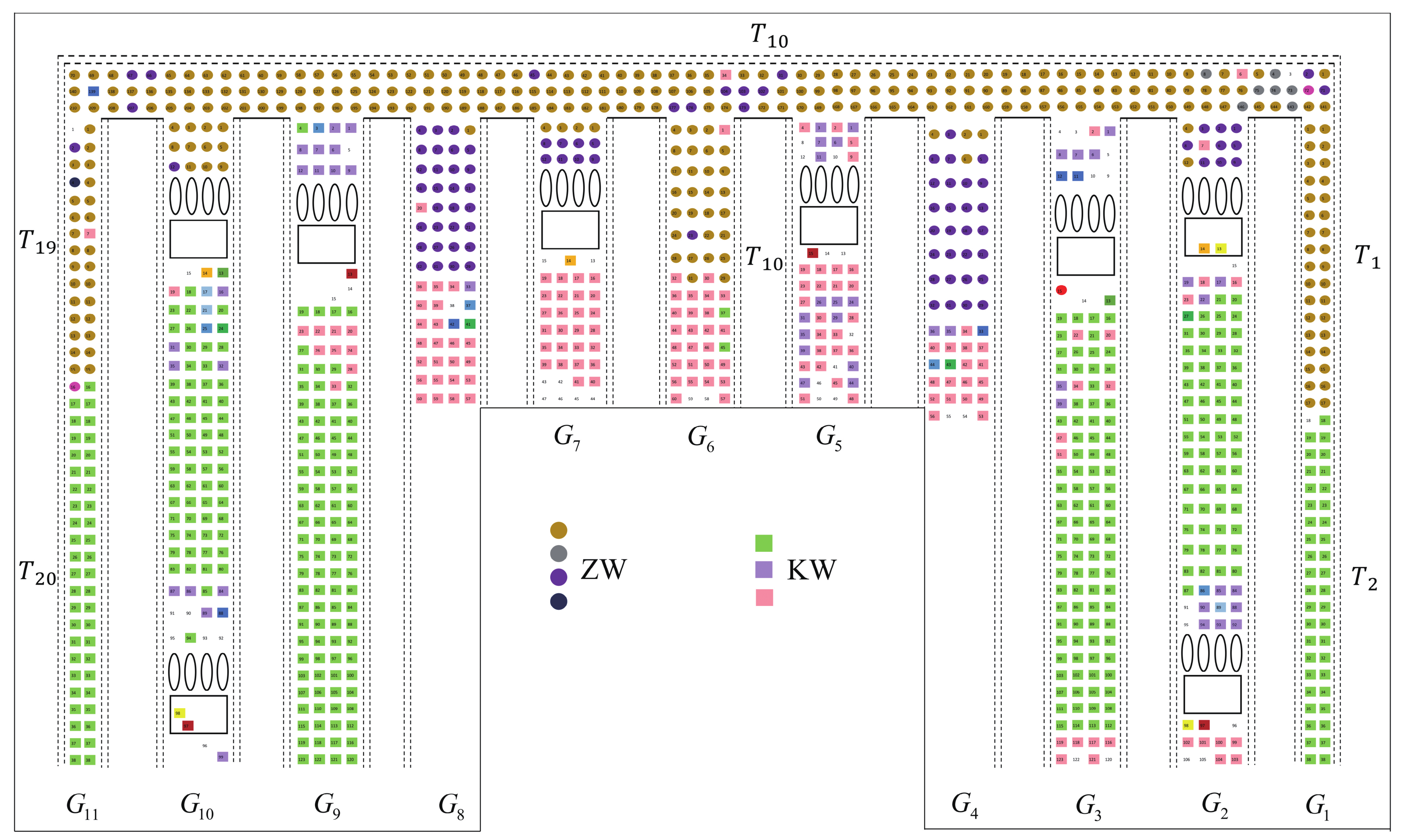}
    \captionsetup{justification=centering}
    \captionof{figure}{The military formation in Pit No. 1.}
    \label{PitNo.1}
\end{center}

\end{document}